\documentclass[final]{IEEEtran}
\usepackage{amsmath,amsfonts}
\usepackage{textcomp}

\usepackage{bm}
\makeatletter

\usepackage{graphicx}
\usepackage[caption=false,font=normalsize,labelfont=sf,textfont=sf]{subfig}
\usepackage{textcomp}
\usepackage{balance}
\usepackage{algorithm}
\usepackage[noend]{algpseudocode}
\usepackage{amssymb}
\usepackage{subcaption}
\usepackage{cite}

\DeclareMathOperator{\sinc}{sinc}

\begin{document}
\title{Real-Time Doppler and Ionospheric Distortion Correction Techniques for Arbitrary Waveforms Utilizing GPU Compute}
\author{Daniel J. Vickers, A. H. Mack, and Idahosa A. Osaretin}

\markboth{}
{}

\maketitle

\begin{figure}[b!]
    \centering
    \begin{minipage}[b][][b]{\linewidth}
        \small{
            DISTRIBUTION STATEMENT A. Approved for public release. Distribution is unlimited.

            This material is based upon work supported by the Department of the Army under Air Force Contract No. FA8702-15-D-0001. Any opinions, findings, conclusions or recommendations expressed in this material are those of the author(s) and do not necessarily reflect the views of the Department of the Army.

            Daniel J. Vickers is with MIT Lincoln Laboratory, Lexington, MA 02421 USA (e-mail: Daniel.Vickers@ll.mit.edu).

            A. H. Mack is with MIT Lincoln Laboratory, Lexington, MA 02421 USA (e-mail: Andrew.Mack@ll.mit.edu). 
            
            Idahosa A. Osaretin is with MIT Lincoln Laboratory, Lexington, MA 02421 USA (e-mail: idahosa.osaretin@ll.mit.edu). 

        }
    \end{minipage}
\end{figure}

\begin{abstract}
    A general requirement for digital signal processing is ionospheric distortion and Doppler dispersion correction, which has historically required radar-specific hardware to implement in real time. With improvements in modern general compute systems, real time digital signal processing is becoming more realizable using non-radar-specific high performance compute. In this paper, we present an analysis of general Doppler correction algorithms for arbitrary waveforms and a novel ionospheric correction algorithm for radar digital signal processing. We also include considerations for efficient implementation of these algorithms in software, specifically using GPU hardware. 
\end{abstract}

\begin{IEEEkeywords}
    Digital radar, GPU, signal processing, arbitrary waveform, Doppler, ionosphere.
\end{IEEEkeywords}

\section{Introduction}

\IEEEPARstart{D}{oppler} and ionospheric dispersion generate significant distortions for ground-space radar and communication systems. Failure to properly correct for these distortions can result in a decrease in the signal-to-noise ratio (SNR) of a received signal, a loss in timing/location accuracy, and more. Doppler time dilation and ionospheric distortion correction can be solved analytically for specific waveforms~\cite{halpin1990propagation}, but general solutions for arbitrary waveforms are computationally expensive and require flexible compute systems. Radar signal processing has historically been implemented in radar-specific hardware such as field-programmable gate arrays (FPGAs), analog components, and application-specific integrated circuits (ASICs) due to real-time processing requirements. While radar-specific hardware provides optimal performance, it often requires longer development times, higher costs, and less-flexible system designs. Software flexibility improvements can be obtained using more-general compute systems like central processing units (CPUs) and graphics processing units (GPUs), which could allow for more-sphistocated digital signal processing (DSP) solutions. The growth trends of computational power for both CPUs and GPUs has been observed~\cite{sun2019summarizing}. As on-chip processing power has improved, software-defined radios (SDRs) leveraging general compute systems have become more common in research, communication, astronomy, defense, and hobbyist use~\cite{henchiri2017digital, javidi2017application, gancio2020upgraded}.

In this paper, we present ionosphere and Doppler dispersion correction methods for use in radar DSP to be performed using non-radar-specific compute for arbitrary waveforms. We also also explore computational optimization of these algorithms for SDR on a GPU for use in real-time processing. Documentation of optimizations and their performance on an NVIDIA H100 GPU are presented here.

Doppler dispersion of a waveform is exactly modeled by the two-way longitudinal relativistic Doppler shift where the frequency, $f$, is dilated such that

\begin{equation}
    f_{doppler} = \frac{f}{\alpha}.
    \label{eq:timedilationeffect}
\end{equation}

Here $\alpha =\frac{1+v_{r}/c}{1-v_{r}/c}$, $v_{r}$ is the range rate of the target, and $c$ is the speed of light~\cite{hamilton1978uniformly}. Note here that we use the convention that $v_{r}$ is positive when the target is approaching the sensor. For a given signal, $S(t)$, the doppler dispersion distorted waveform is given as:

\begin{equation}
    S_{doppler}(t) = S(\alpha t).
    \label{eq:timedilatedsignal}
\end{equation}

For any signals with a known analytic representation, (\ref{eq:timedilatedsignal}) can be computed exactly by resampling the signal according to the target range rate. More specifically, when correcting linear frequency modulated waveforms (LFMs), this can be modeled by generating a new LFM with a modified phase slope according to (\ref{eq:timedilationeffect}). This analytic resampling method is typically ideal because of its exact correction and because the time complexity scales as $\mathcal{O}\left(N\right)$. However, when performing DSP, the transmit and receive signals are discretized by sampling the waveform. It can then become desirable to pass these samples between software interfaces in order to simplify the overall software complexity of the system. In these cases where no analytic representation of the signal is available, one is forced to use the general solution of approximating the resampling numerically. There are many known ways to compute this numerical resampling. For this paper, we considered these methods:

\begin{itemize}
    \item Time Domain Downconversion or Frequency Conversion: The signal is multiplied by a tone in the time domain, resulting in an upconversion or downconversion. This algorithm is extremely computationally efficient as it scales with a time complexity of $\mathcal{O}\left(N\right)$. However, the frequency shift is only exact at one frequency, and induces errors for pulses with large bandwidths and/or steep phase slopes~\cite{chen2002time}. For LFMs, this error scales as $\mathcal{O}\left(B^{2}\right)$, where $B$ is the bandwidth of the signal. For arbitrary waveforms, one typically expects this error to be more-severe.

    \item Polynomial Interpolation: A linear, cubic, higher order, or spline interpolation function is used. While this method is computationally efficient, typically scaling as $\mathcal{O}\left(N\right)$, it results in a loss of power in the signal, especially at turnovers in the value. Higher sample rate signals can result in less error, but raise the computational load.

    \item FFT P/Q Resampling: The Fourier transform of the signal has terms removed or padded and then the inverse Fourier transform is taken. This results in the exact resampling assuming a Nyquist limited waveform; however, this method can be computationally expensive due to requiring discrete Fourier transforms (DFTs) of non-ideal numbers of samples. This leads to variable sample rates that depend upon the specific number of sample used, and scales as $\mathcal{O}\left(N\log{N}\right)$ in the ideal case. This method is also only exact out to a integer number of samples to pad or remove. A brief exploration of the accuracy and performance of FFT resampling can be found in Ref.~\cite{fraser1989interpolation}.

    \item Sinc Interpolation: Allows exact resampling of Nyquist limited signals. The exact form of sinc interpolation grows as $\mathcal{O}\left(N^{2}\right)$ in time, but windowing the filter to a size of $N_{window}$ samples results in a time complexity of $\mathcal{O}\left(N\times N_{window}\right)$. The choice of taper window size, taper, initial sampling rate, and other features determines the performance of this algorithm. An overview of Whittaker-Shannon interpolation and general interpolation of Nyquist limited signals can be found in Ref.~\cite{marks2012introduction}.
\end{itemize}

Ionosphere distortion correction for waveforms has been performed previously with various methods. One high-accuracy method approximates solutions calculated for specific waveforms~\cite{halpin1990propagation}. For the same reason as Doppler dispersion correction, this analytic method can be ideal for cases where known simple waveforms will be transmitted. However, the choice to use an analytic method typically increases system complexity and reduces system flexibility. For numerical approximations, another class of  methods involves using finite impulse response (FIR) filters with a frequency dependent response matched to the ionosphere~\cite{oppenheim1999discrete, proakis2007digital, Parks1972Chebyshev}. An overview of modern FIR filtering methods can be found in Ref.~\cite{lehtinen2019optimal}. These FIR filters typically operate by representing the ionospheric distortion as a frequency response with three unique input coefficients. All but one of the coefficients are held constant while the variable coefficient is tuned to optimal performance. These methods are computationally efficient, but are not exact and are commonly best for applications where one or more parameters of the frequency response are unknown.

In this paper, we will present an ionosphere distortion correction method that is performed computationally and is valid for arbitrary waveforms. We will also present an analysis of various Doppler dispersion correction methods via interpolation for arbitrary waveforms. We will be leveraging modern hardware to explore the implementation of these algorithms in real time on a GPU as well as provide an documentation of various optimization strategies considered for implementation of these algorithms.

\section{Ionospheric Distortion Correction}

A review of ionospheric effects on radio signals can be found in Appendix~\ref{app:ionospherereview}. In brief, ionospheric distortion of a waveform is modeled as a frequency-based time delay of the signal, given as

\begin{equation}
    \begin{split}
    \tau(f) = \frac{e^2}{8\pi^{2} m_{e} \epsilon_{0} cf^2} E = \frac{K_{2}}{cf^{2}}.
    \end{split}
    \label{eq:ionosphericdelayequation}
\end{equation}

In (\ref{eq:ionosphericdelayequation}), $E$ is the total electron content (TEC) along the path of propagation and $K_{2}$ is the plasma delay constant.

To correct for the ionospheric distortion of a waveform, an inverse time shift can be applied by temporally translating the waveform $-2\tau(f)$ (to account for the two-way propagation). For continuous waveforms (CWs) with constant frequency, ionospheric effects are reduced to a uniform translation in time (or phase). This problem becomes non-trivial when correcting for waveforms with frequency modulation and more complex frequency characteristics. To perform real-time correction of ionospheric effects for receive waveforms, we sought a simplified approach to previously used methods. By leveraging modern hardware to perform more computationally intense algorithms, our goal was a more-general correction method that results in an overall simpler algorithm and software interface. In this section, we present a mathematic background of the ionospheric correction methods. We then present our numeric method for correction ionospheric distortion and compare this numeric method to analytic methods. Finally, we discuss the time performance of this method.

\subsection{Ionospheric Dispersion Compensation Via Predistortion}

A known method to analytically compensate for ionospheric dispersion is to apply a correction via predistortion, where the transmitted waveform is inversely distorted in relation to the predicted ionospheric effect. The goal being to compensate for the ionosphere such that the desired waveform is received. This method requires precise analytic solutions or approximations for any given frequency-modulated waveform. One of these predistortion methods was developed for LFMs, and was presented by Halpin, Urkowitz, and Maron~\cite{halpin1990propagation}. The derivation of their polynomial approximation can be found in Appendices A and C of Ref.~\cite{halpin1990propagation}. Their result is repeated here in (\ref{eq:halpintimedependenceinversion}):

\begin{equation}
    f^3 - \left(f_{0} + \frac{Bt}{T} - \frac{2K_{2} B}{cf_{0}^{2} T}\right) f^{2} - \frac{2K_{2} B}{cT} = 0.
    \label{eq:halpintimedependenceinversion}
\end{equation}

Here $T$, $B$, and $f_{0}$ are the pulse width, bandwidth, and start frequency of the LFM, respectively.\ (\ref{eq:halpintimedependenceinversion}) has typically been solved by approximating the frequency as a polynomial expansion of the form

\begin{equation}
    f(t) = f_{0} + (\mu_{0} + \Delta\mu)t + \gamma t^{2}.
    \label{eq:ionospherefrequencypolynomialexpansion}
\end{equation}

There are many ways to determine the coefficients of (\ref{eq:ionospherefrequencypolynomialexpansion}). Here we will note the Pseudo-Chebyshev polynomial expansion method\footnote{Note that the results presented in (\ref{eq:T1}) and (\ref{eq:T2}) differ from those presented in Ref.~\cite{halpin1990propagation} by a factor of 2 on two terms. We concluded that this factor of 2 difference is a minor error and elaborate on this result in Appendix~\ref{app:k2correction}.} from Ref.~\cite{halpin1990propagation} in which the coefficients $\mu_{0}$, $\Delta\mu$, and $\gamma$ are calculated as

\begin{equation}
    \mu_{0} = \frac{B}{T},
    \label{eq:mu0}
\end{equation}

\begin{equation}
    \Delta\mu = \frac{aB}{2} \frac{ T_{1}^{2} + T_{3}^{2} }{T_{1} T_{2} \left( T_{3} - T_{1} \right)} - \mu_0,
    \label{eq:deltamu}
\end{equation}

and

\begin{equation}
    \gamma = \frac{aB}{2} \frac{ T_{2}}{T_{1} T_{3} \left( T_{3} - T_{1} \right)},
    \label{eq:gamma}
\end{equation}

where $a=\sqrt{3}/2$ and the coefficients $T_{1}$ and $T_{2}$ are given by

\begin{equation}
    \begin{split}
    T_{1} = -\frac{aT}{2} &+ \frac{2K_2}{c} \bigg[ \left(f_{0} + \frac{B}{2}\right)^{-2} \\
    & - \left(f_{0} + \frac{B(1-a)}{2} \right)^{-2} \bigg],
    \end{split}
    \label{eq:T1}
\end{equation}

and

\begin{equation}
    \begin{split}
    T_{2} = \frac{2K_2}{c} & \bigg[2\left(f_{0} + \frac{B}{2} \right)^{-2} - \left(f_{0} + \frac{B(1-a)}{2} \right)^{-2}\\
    & - \left(f_{0} + \frac{B(1+a)}{2} \right)^{-2} \bigg].
    \end{split}
    \label{eq:T2}
\end{equation}

We also solved (\ref{eq:halpintimedependenceinversion}) directly via the cubic formula to yield

\begin{equation}
    f(t) = \frac{1}{3}\left( f_{0} + \frac{Bt}{T} - \frac{2K_{2}B}{cf_{0}^{2}T} - \omega - \frac{\chi}{\omega} \right).
    \label{eq:frequencycubicequationresult}
\end{equation}

Where the coefficients are given by

\begin{equation}
    \chi = \left(f_{0} + \frac{Bt}{T} - \frac{2K_{2}B}{cf_{0}^{2}T}\right)^{2},
    \label{eq:chi}
\end{equation}

\begin{equation}
    \psi = -2\left( f_{0} + \frac{Bt}{T} - \frac{2K_{2}B}{cf_{0}^{2}T} \right)^{3} -27\frac{2K_{2}B}{cT},
    \label{eq:psi}
\end{equation}

and

\begin{equation}
    \omega = \sqrt[3]{\frac{\psi + \sqrt{\psi^{2} - 4 \chi^{3}}}{2}}.
    \label{eq:omega}
\end{equation}

Note that (\ref{eq:halpintimedependenceinversion}) can have up to three unique solutions. For a practical choice of parameters, (\ref{eq:frequencycubicequationresult}) is the only purely real solution available. The only physical solutions for frequency are purely real, and therefore the only solution considered is that shown in (\ref{eq:frequencycubicequationresult}).

Recall that we can represent a phase modulated waveform as $s(t) = \exp[i\phi(t)]$, where $\phi(t)=2\pi\int_{0}^{t}f(t')dt'$. The case of (\ref{eq:frequencycubicequationresult}) is too complicated to integrate directly and yield an analytic representation of the predistorted waveform's phase. However, the phase can be integrated numerically to provide another approximation method for correction of ionospheric effects via predistortion with error related to the method of numerical integration. Both of the predistortion methods shown in (\ref{eq:ionospherefrequencypolynomialexpansion}) and (\ref{eq:frequencycubicequationresult}) provide useful points of comparison for accuracy of ionospheric correction methods.

\subsection{Ionospheric Dispersion Compensation Via FFT Phase Correction}

In this section, we propose a method for numerically determining the ionospheric distortion correction of a transmitted or received waveform in real time directly from discrete samples. Recall that given a function $g(t)$, we represent the Fourier transform as $\hat{g}(f)=\int{g(t)\exp(-2\pi itf)dt}=\mathcal{F}(g(t))$. Note that for any function, if $g(t-t_0)=h(t)$ then $\hat{g}(f)\exp(-2\pi it_{0}f)=\hat{h}(f)$. If $t_0$ is a constant, then $\exp(-2\pi it_{0}f)$ is a simple tone, and $h(t)$ is a linear translation by time $t_0$. If instead, we make $t_0$ a function of frequency, we can translate each frequency in our function by a frequency dependent amount. Using the temporal translation from (\ref{eq:ionosphericdelayequation}) we can conclude that the ionospheric effect on a transmitted waveform, $s_{trans}$, due to two-way propagation will result in a received waveform, $s_{rec}$, of

\begin{equation}
    \begin{split}
    s_{rec} &= \mathcal{F}^{-1} \left[ \mathcal{F}(s_{trans}) \exp\left(2\pi i(2\tau)f\right) \right] \\
    &= \mathcal{F}^{-1} \left[ \mathcal{F}(s_{trans}) \exp\left( \frac{4\pi iK_{2}}{cf} \right) \right].
    \end{split}
    \label{eq:ionospheredistortionmethod}
\end{equation}

On its own, (\ref{eq:ionospheredistortionmethod}) is a convenient algorithm to directly apply ionospheric distortion to arbitrary waveforms in software. Solving for the inverse of (\ref{eq:ionospheredistortionmethod}) results in a method to remove the ionospheric distortion accrued during propagation and is identical to applying a temporal shift of $-2\tau(f)$. The inversion of (\ref{eq:ionospheredistortionmethod}) yields

\begin{equation}
    \begin{split}
    s_{trans} &= \mathcal{F}^{-1} \left[ \mathcal{F}(s_{rec}) \exp\left(2\pi i(-2\tau)f\right) \right] \\
    &= \mathcal{F}^{-1} \left[ \mathcal{F}(s_{rec}) \exp\left(\frac{-4\pi iK_{2}}{cf} \right) \right].
    \end{split}
    \label{eq:ionospherecorrectionmethod}
\end{equation}

A similar method to (\ref{eq:ionospherecorrectionmethod}) was applied by Peters, Schroeder, and Romero-Wolf~\cite{peters2020passive} for post-processing of satellite data when determining the TEC values of Europa's atmosphere. Here, we intend for the algorithm to correct for the ionosphere in real time given a known TEC value.

In practice, the Fourier transformations used in (\ref{eq:ionospherecorrectionmethod}) must be done discretely as Fast Fourier Transforms (FFTs) and inverse FFTs (iFFTs). Performing the FFT and iFFT for each received waveform has been too slow to solve in the past for real time applications on high-bandwidth signals. Modern improvements to hardware and algorithms have made this computation achievable to implement digitally in real time for high bandwidth signals.

Compared to using a polynomial expansion technique, our proposed method offers several benefits. The technique shown in (\ref{eq:ionospherecorrectionmethod}) is a direct calculation of the frequency dependent phase shift. This results in a few lines of code, allowing a first step in prototyping with little chance for errors. Our technique can distort a waveform on transmit (predistortion) or on receive (post-distortion), and is valid for arbitrary waveforms. The algorithm's only requirement is that the waveform is sampled within the Nyquist limit. When an analytical solution is inaccessible, (\ref{eq:ionospherecorrectionmethod}) can be used as a general numerical method.

\begin{figure}[h]
    \centering
    \includegraphics[clip, trim=2em 16em 2em 11em, width=\linewidth]{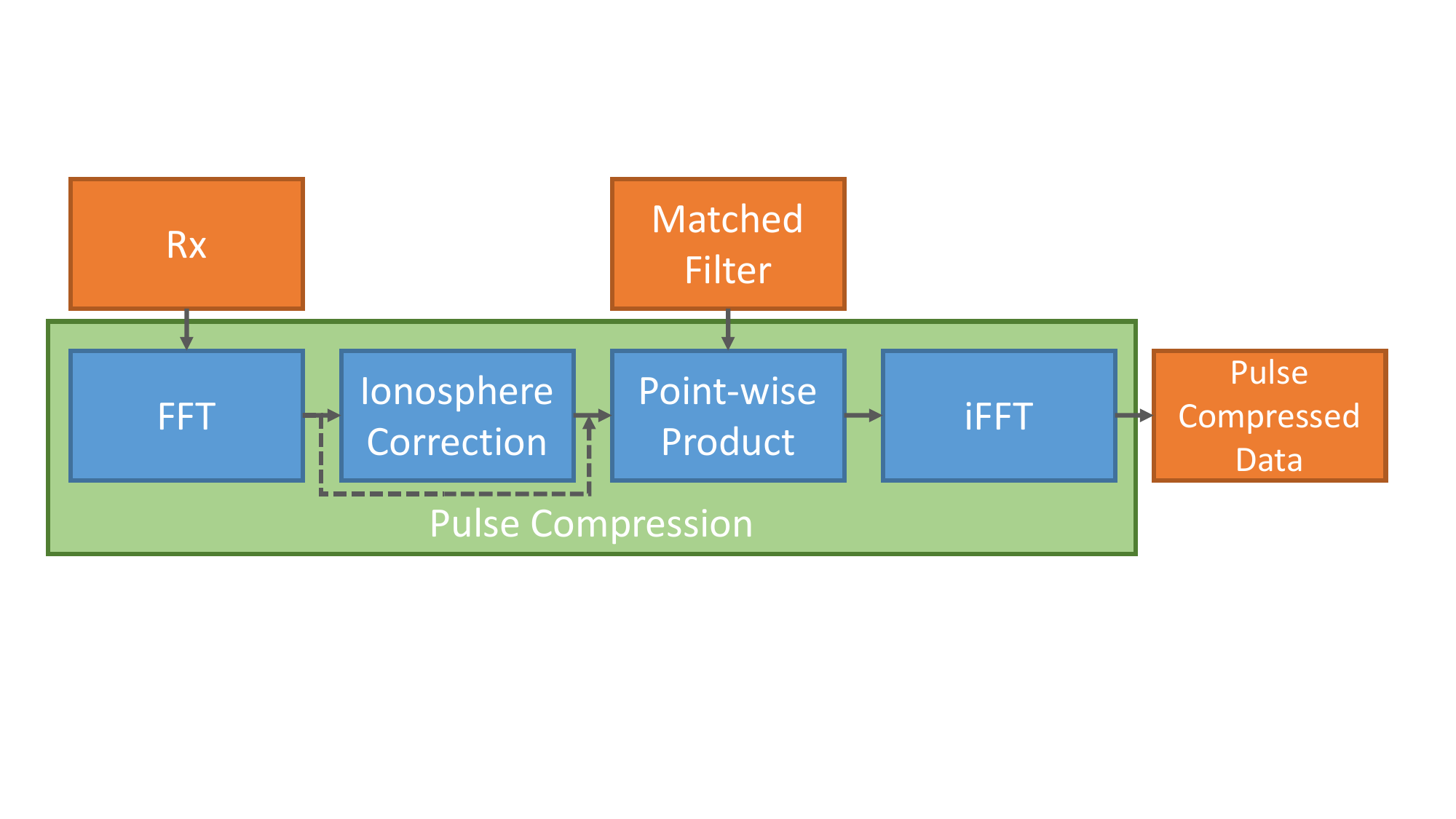}
    \caption{Overview of process of performing pulse compression on the GPU with ionospheric distortion correction.}
    \label{fig:pulse_compression_path}
  \end{figure}

The primary drawback of our proposed method is that it is more computationally expensive in cases where a polynomial expansion method can be used. This high computational cost is primarily due to the use of the FFT and iFFT algorithms. This cost can be mitigated with the use of modern hardware. The algorithm shown in (\ref{eq:ionospherecorrectionmethod}) is also an excellent candidate for parallelization on a GPU. The computational work done in the frequency domain can also be bundled with other calculations that require FFTs in the signal processing chain and benefit from the efficiency of the shared use of FFT compute time. An example flow diagram of how one could combine pulse compression with (\ref{eq:ionospherecorrectionmethod}) for post-distortion is shown in Fig.~\ref{fig:pulse_compression_path} These optimizations can make our method viable in SDRs to be used for real-time processing in radars and communication systems.

\subsection{Comparison to Ionospheric Correction Predistortion Methods}

\begin{figure}[h]
    \centering
    \includegraphics[width=1\linewidth]{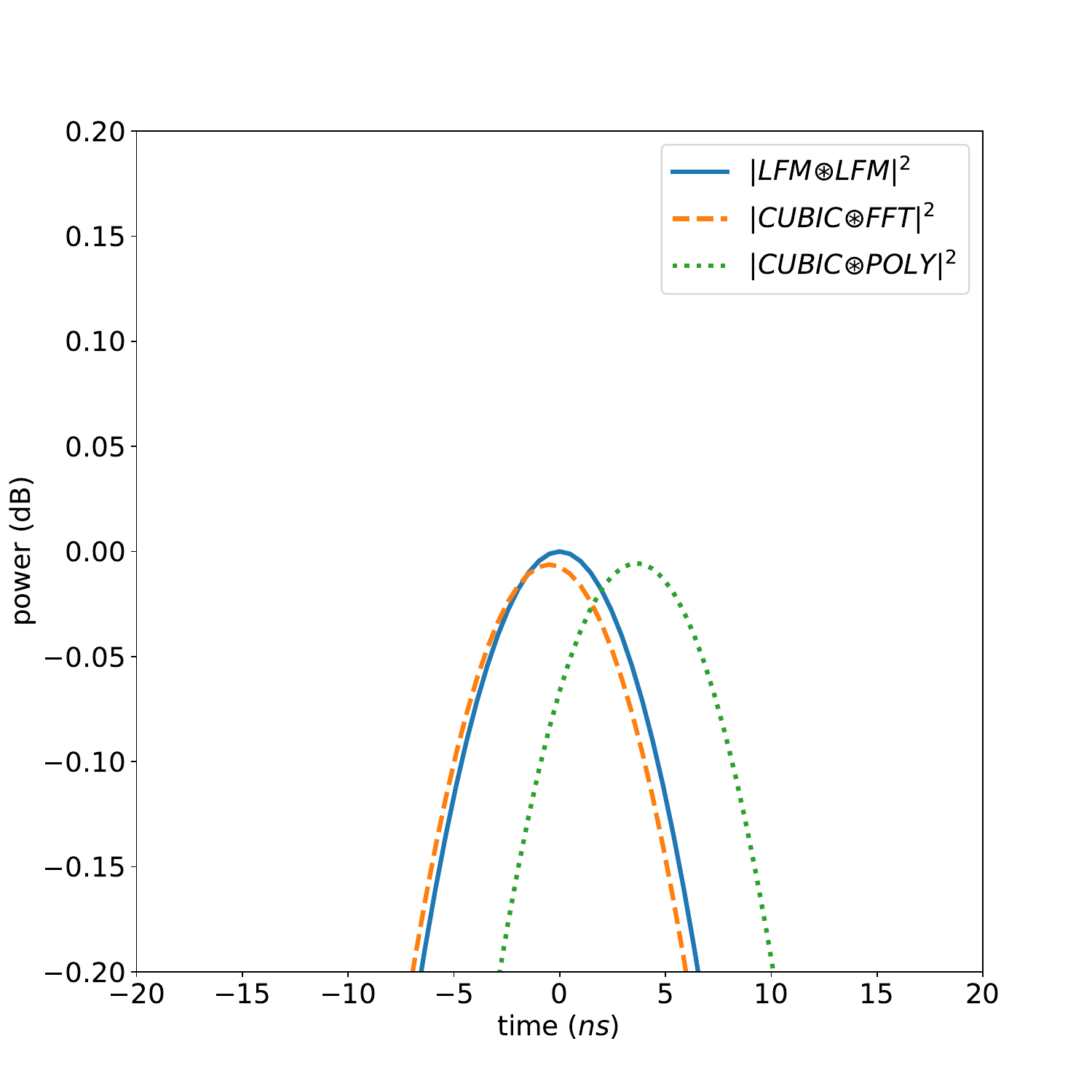}
    \caption{Comparison of our FFT-based method with both the undistorted waveform (labeled LFM), the Pseudo-Chebyshev polynomial approximation (labeled POLY), our FFT based method (labeled FFT), and our exact frequency from the cubic formula shown in Eq.~\ref{eq:frequencycubicequationresult} integrated via trapezoidal integration (labeled CUBIC). This test was run with $f_{0} = 413~\text{MHz}$, $B = 18~\text{MHz}$, a sample rate of $f_{s} = 2.048~\text{GHz}$, $E = 100e16~\text{electrons}/\text{m}^{2}$, and $T=100~\mu \text{s}$. As we can see, compared to the total amount of energy in our original signal, shown in the $\text{LFM}\circledast\text{LFM}$ curve, we lose little SNR for the $\text{CUBIC}\circledast\text{POLY}$ curve, which is the cross correlation the CUBIC waveform with the POLY approximated waveform. In this particular example, the $\text{CUBIC}\circledast\text{POLY}$ curve lost $<0.01~\text{dB}$ SNR. The $\text{CUBIC}\circledast\text{FFT}$ curve compares the CUBIC waveform with our FFT waveform method. For this example, the $\text{CUBIC}\circledast\text{FFT}$ compared similarly to the $\text{CUBIC}\circledast\text{POLY}$ curve as close as $<0.001~\text{dB}$, indicating that our FFT method is on par with the accuracy of the Pseudo-Chebyshev polynomial method.}
    \label{fig:pseudochebyshevcomparison}
\end{figure}

We tested our method against results for LFM predistortion using the Pseudo-Chebyshev approximation method and the cubic solution from (\ref{eq:frequencycubicequationresult}) using trapezoidal integration to determine the phase. We chose the Pseudo-Chebyshev method because it generally has one of the highest SNRs for LFM predistortion methods. Results from one of our sample test cases can be seen in Fig.~\ref{fig:pseudochebyshevcomparison}.

As can be seen in this example, the FFT approximation method and the Pseudo-Chebyshev method resulted in losses on the order of $<0.01~\text{dB}$ SNR or less when compared to the cubic solution. The difference between each method resulted in peaks separated by $<0.001~\text{dB}$. This analysis indicated that our FFT method is approximately as accurate as the Pseudo-Chebyshev polynomial approximation method for the parameters used to generate Fig.~\ref{fig:pseudochebyshevcomparison}.

\subsection{Ionospheric Correction Implementation and Benchmarking}

\begin{figure*}[t]
    \centering
    \subfloat[]{
    \begin{minipage}[t]{0.49\textwidth}
        \centering
        \includegraphics[width=1\linewidth]{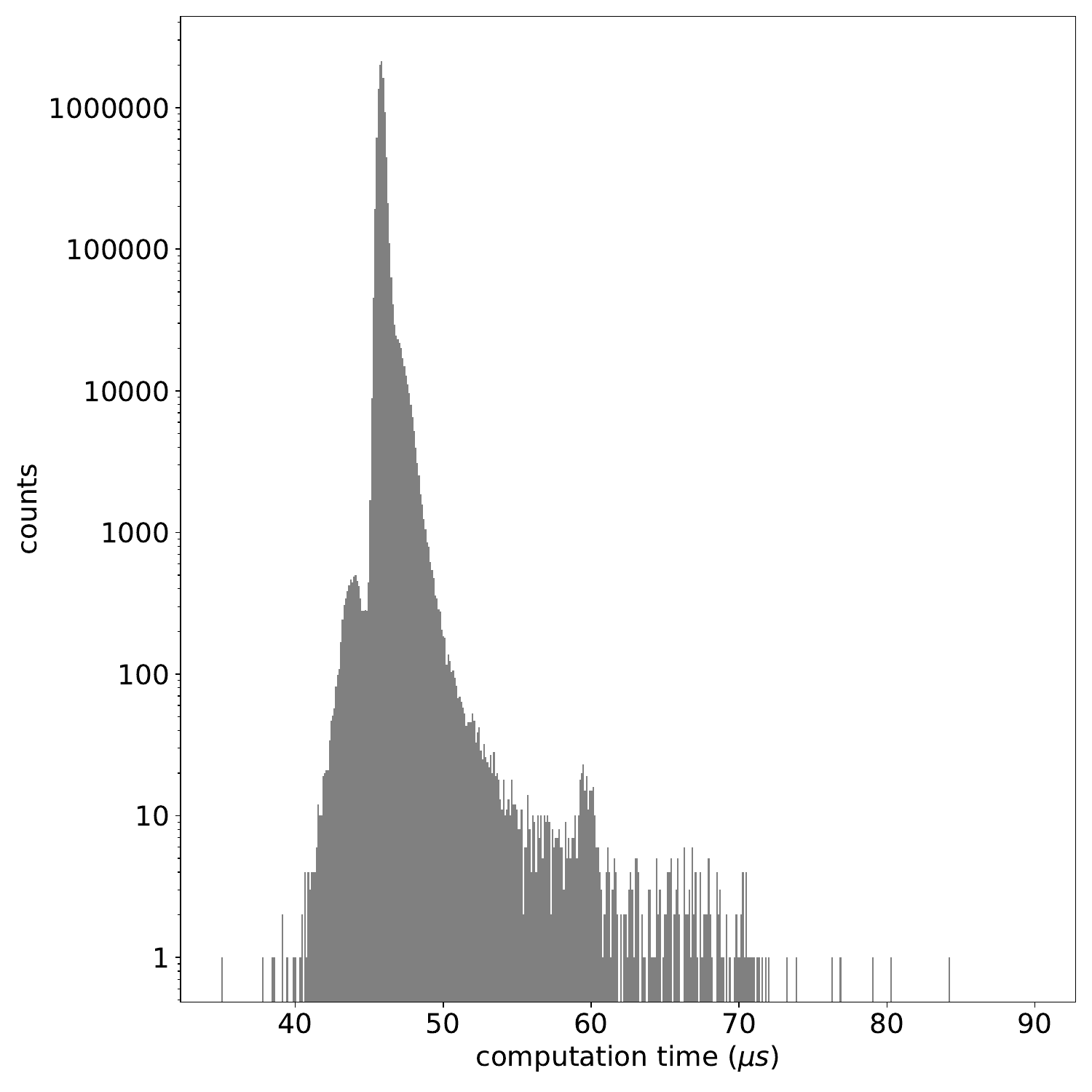}
        \label{fig:ionobenchmarkwcompression}
    \end{minipage}}
    \hfill
    \subfloat[]{
    \begin{minipage}[t]{0.49\textwidth}
        \centering
        \includegraphics[width=1\linewidth]{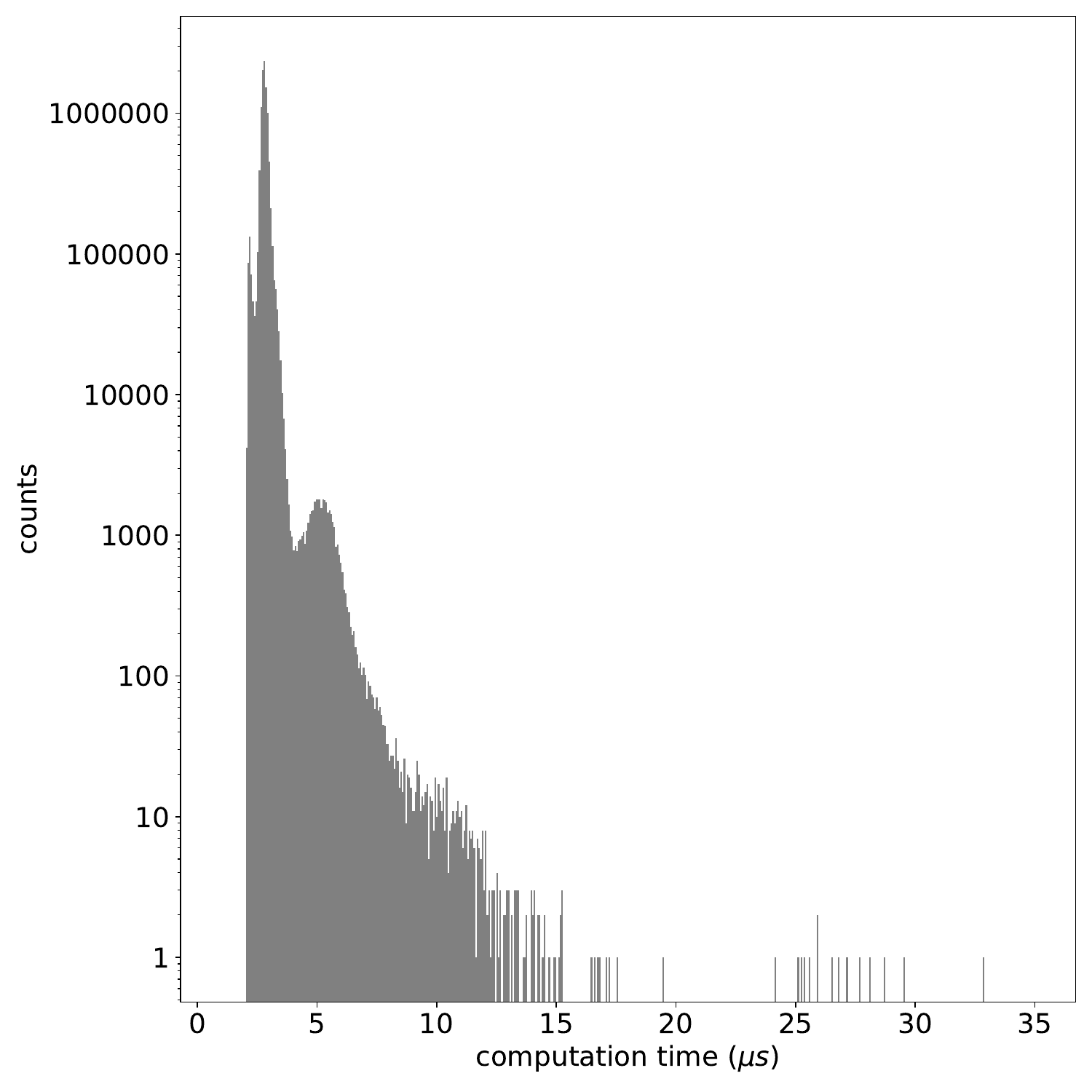}
        \label{fig:ionobenchmarkwocompression}
    \end{minipage}}
    \hfill
    \caption{Histograms showing the results from $10^{7}$ time trials where we performed the algorithm shown in (\ref{eq:ionospherecorrectionmethod}). This test was run with $f_{0} = 413~\text{MHz}$, $B = 18~\text{MHz}$, a sample rate of $f_{s} = 2.048~\text{GHz}$, $E = 100e16~\text{electrons}/\text{m}^{2}$, $T=100~\mu \text{s}$, and had $2^{19}$ total I/Q samples. Figure~\ref{fig:ionobenchmarkwcompression} shows the rates when performing ionospheric correction during pulse compression, which averaged about $46~\mu\text{s}$ to complete and correlates to a rate of approximately $11.3~\text{Gigasamples}/\text{s}$. Figure~\ref{fig:ionobenchmarkwocompression} excludes the FFT and iFFT times and averages approximately $2200~\text{ns}$, which correlates to about $240~\text{Gigasamples}/\text{s}$.}
    \label{fig:ionobenchmark}
\end{figure*}

To demonstrate the computational rates of the methods considered, all tests in this paper were implemented in CUDA 12.3 and performed on the NVIDIA H100 GPU with a $3.25~\text{GHz}$ CPU processor connected via PCIe 5.0. This choice of hardware allowed us to parallelize the FFT, iFFT, and the phase shifting of each sample in our receive window. Figure~\ref{fig:ionobenchmark} shows the benchmarking result of running (\ref{eq:ionospherecorrectionmethod}) on a realistic waveform. Notably Fig.~\ref{fig:ionobenchmarkwocompression} shows benchmarks excluding the FFT and iFFT compute times.

%

As can be seen in Fig.~\ref{fig:ionobenchmark}, on our hardware we were able to complete the ionospheric correction considered in about $46~\mu\text{s}$; nearly $44~\mu\text{s}$ of which are from performing the required DFTs. We separate these test cases to demonstrate the efficiency that can be gained from bundling the ionospheric correction with other digital signal processing methods that require computations in the frequency domain. Common digital signal processing methods, including pulse compression, often require DFTs to be performed, and the ionospheric correction can be run in tandem with these other algorithms to allow for optimal signal processing rates. This efficiency makes our ionospheric correction method viable for use in digital processing in real time.

\section{Doppler Dispersion Correction}

Similar to our approach to ionospheric distortion correction, we also investigated numeric Doppler dispersion correction methods on modern hardware. Because Doppler dispersion is already well-modeled analytically, this section will be focused on novel implementation of these numeric interpolation algorithms on the GPU. We first start with an overview of all relevant methods based on their time complexity and accuracy. We then provide an more-thorough analysis of the time performance and accuracy of one of the most-promising numerical methods that we present.

\subsection{Doppler Dispersion Time Dilation Methods}

In this section, we present an overview analysis of various Doppler-dispersion correction methods, all of which seek to approximate the effect of (\ref{eq:timedilatedsignal}). Most methods will approximate (\ref{eq:timedilatedsignal}) via a numeric interpolation of the transmitted signal. Table~\ref{tab:dopplercomparison} lists the Doppler dispersion correction methods considered here along with some basic features of each method\footnote{We determined the error drop off for windowed Whittaker-Shannon interpolation as well as frequency conversion numerically. The data presented in the table is reflective of that drop off for LFMs.}.\\ 

Whittaker-Shannon interpolation (also known as sinc interpolation) is an exact interpolation method for Nyquist limited signals, and is given by (\ref{eq:whittakershannoninterpolation}):

\begin{equation}
    x(t)=\sum_{n=-\infty}^{\infty} x[n] \sinc\left(\frac{t-n\Delta t}{\Delta t}\right).
    \label{eq:whittakershannoninterpolation}
\end{equation}

\begin{table*}[t]
\caption{Comparison of features of various Doppler dispersion correction/discrete resampling methods.}
\label{tab:dopplercomparison}
\centering
\begin{tabular}{ ||c c c c|| }
    \hline
    Method Name & Time Complexity & Accuracy & Arbitrary Waveforms \\
    \hline\hline
    Whittaker-Shannon Interpolation & $\mathcal{O}\left( N^{2} \right)$ & Exact & Yes \\
    \hline
    Windowed Whittaker-Shannon Interpolation & $\mathcal{O}\left( N\times N_{window} \right)$ & $\mathcal{O}\left( N_{window}^{-2} \right)$ & Yes \\
    \hline
    Frequency Conversion & $\mathcal{O}\left( N \right)$ & $\mathcal{O}\left( B^{2} \right)$ (for LFMs) & Yes \\
    \hline
    FFT P/Q Resampling & Variable & Exact for Even Whole Indices & Yes \\
    \hline
    FFT P/Q Resampling w/ Frequency Conversion & Variable & Exact for Even Whole Indices & Yes \\
    \hline
    Analytical Resampling & $\mathcal{O}\left( N \right)$ & Exact & No \\
    \hline
    Linear Interpolation & $\mathcal{O}\left( N \right)$ & $\mathcal{O}\left( N^{-1} \right)$ & Yes \\
    \hline
\end{tabular}
\end{table*}

Here, $\Delta t$ is the sample period of the discrete signal. (\ref{eq:whittakershannoninterpolation}) is general and does not require upsampling to achieve exact performance. Unfortunately, the time complexity of Whittaker-Shannon interpolation scales as $\mathcal{O}\left(N^{2}\right)$. For more optimal interpolation rates, Whittaker-Shannon requires a window where only $N_{window}$ neighboring points are used in the summation. This was investigated in Ref.~\cite{Liao2018InterpolationMethods}.

\begin{figure}[h]
    \centering
    \includegraphics[width=1\linewidth]{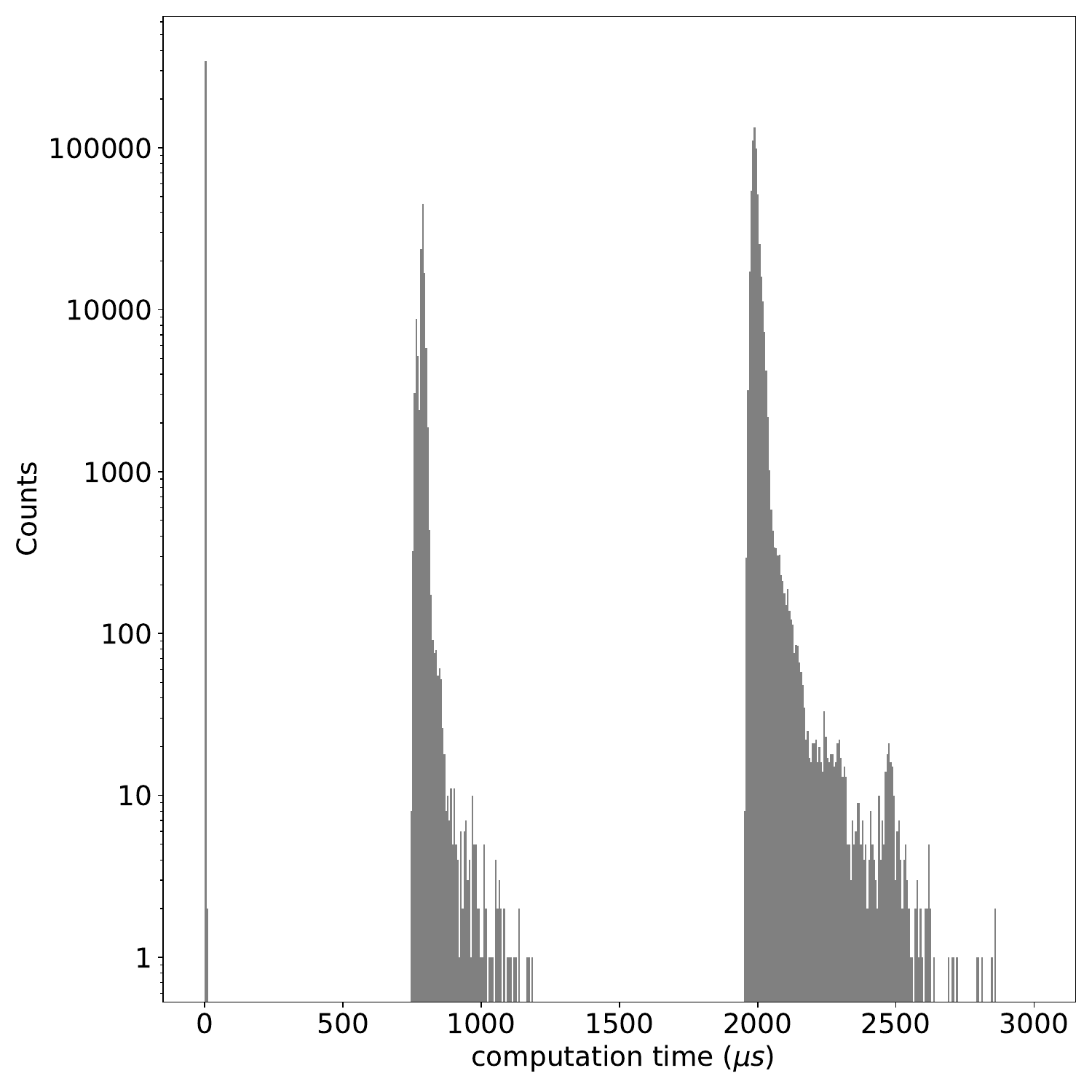}
    \caption{Doppler dispersion correction rates for $10^{6}$ trials of FFT P/Q Resampling with a waveform such that $f_{0}=420~\text{MHz}$, $B=18~\text{MHz}$, and $T=500~\mu\text{s}$. To demonstrate the variable time complexity, we also varied the target velocity by using a uniform distribution between $v=0 \text{ to } 5~\text{km}/\text{s}$. Notice three separate clusters of completion times, each related to a different range of velocities. The left-most cluster is a sharp peak with times $\leq10~\mu\text{s}$ and corresponds to about $v<1500~\text{km}/\text{s}$ where no samples were added (no work was done). The right-most peak was for velocities such that about $1500<v<4500~\text{km}/\text{s}$, where only 2 samples were removed. The final middle cluster was the remaining trials during which 4 samples were removed. The average completion rates for the left, center, and right cluster are about $6.5~\text{Terasamples}/\text{s}$, $670~\text{Megasamples}/\text{s}$, and $260~\text{Megasamples}/\text{s}$ respectively.}
    \label{fig:pqbenchmark}
\end{figure}

For FFT P/Q resampling, the signal can be resampled during an FFT by padding the signal with an even number of samples. When the number of interpolants, $N_{I}$, is related to the number of input samples by a factor of a power of 2, FFT P/Q resampling completes in $\mathcal{O}\left(N_{I}\log{N_{I}}\right)$ time, but the time complexity is much more sophisticated and almost always less optimal in other cases. The variable time complexity is demonstrated in Fig.~\ref{fig:pqbenchmark} where three separate clusters of Doppler dispersion correction rates can be observed, each cluster relating to a range of target velocities. When the desired number of interpolants is not an even integer, FFT P/Q resampling is not exact. FFT P/Q Resampling can be combined with frequency conversion for fine-tune adjustment between even-integer interpolant results. This is done by first performing FFT P/Q Resampling to the nearest velocity that removes an even number of samples, and then performing frequency conversion to shift the remaining frequency difference. This presents a trade space for the optimal time dilation method that depends on the center frequency, bandwidth, signal length, and modulation.

\begin{figure}[h]
    \centering
    \includegraphics[width=1\linewidth]{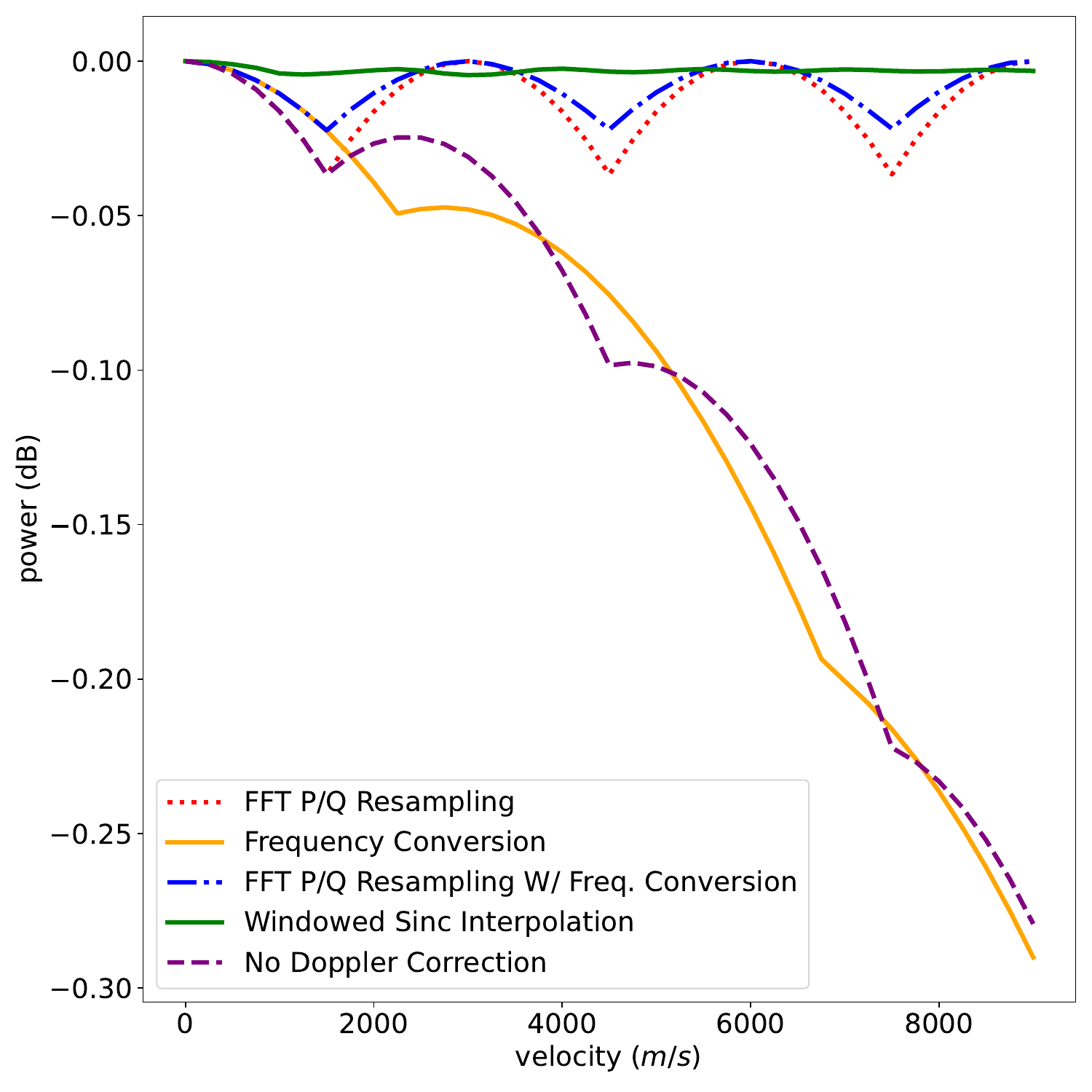}
    \caption{Comparison of error as a function of target velocity for the various Doppler dispersion correction methods considered in Table~\ref{tab:dopplercomparison}. The waveform used here had $f_{0}=411~\text{MHz}$, $B=18~\text{MHz}$, and $T=500~\mu\text{s}$. The waveform was originally sampled at a rate of $f_{s0}=2~\text{GHz}$, and was downconverted to $f_{s}=200~\text{MHz}$. Notice that the the error for FFT P/Q resampling is harmonic, with no error at velocities that require an even integer number of samples be padded and maximal error for velocities that require an odd integer number of samples be added. Note that for the sinc interpolation loss shown here, a window size of $N_{window}=25~\text{samples}$ was used as well as a lookup table (LUT) that was sampled such that for $\sinc (x)$, $\Delta x=1/100$ ($100~\text{values}/\text{sample}$). Refer to Appendix~\ref{app:sincoptimizations} for details on the implementation of LUTs.}
    \label{fig:dopplersnrlosscomparison}
\end{figure}

An overview of the accuracy of each method is summarized in Fig.~\ref{fig:dopplersnrlosscomparison}. For small dilations, the frequency conversion method has the lowest computational load and performs well. Even when at even sample removal, the FFT P/Q resampling with or without the tone has nearly equivalent accuracy to Whittaker-Shannon interpolation at the cost of variable compute times. At non-integer interpolants, the sinc/Whittaker-Shannon method performs the dilation with the least error.

\subsection{Implementation of Whittaker-Shannon Interpolation}

\begin{figure}[h]
    \centering
    \includegraphics[width=1\linewidth]{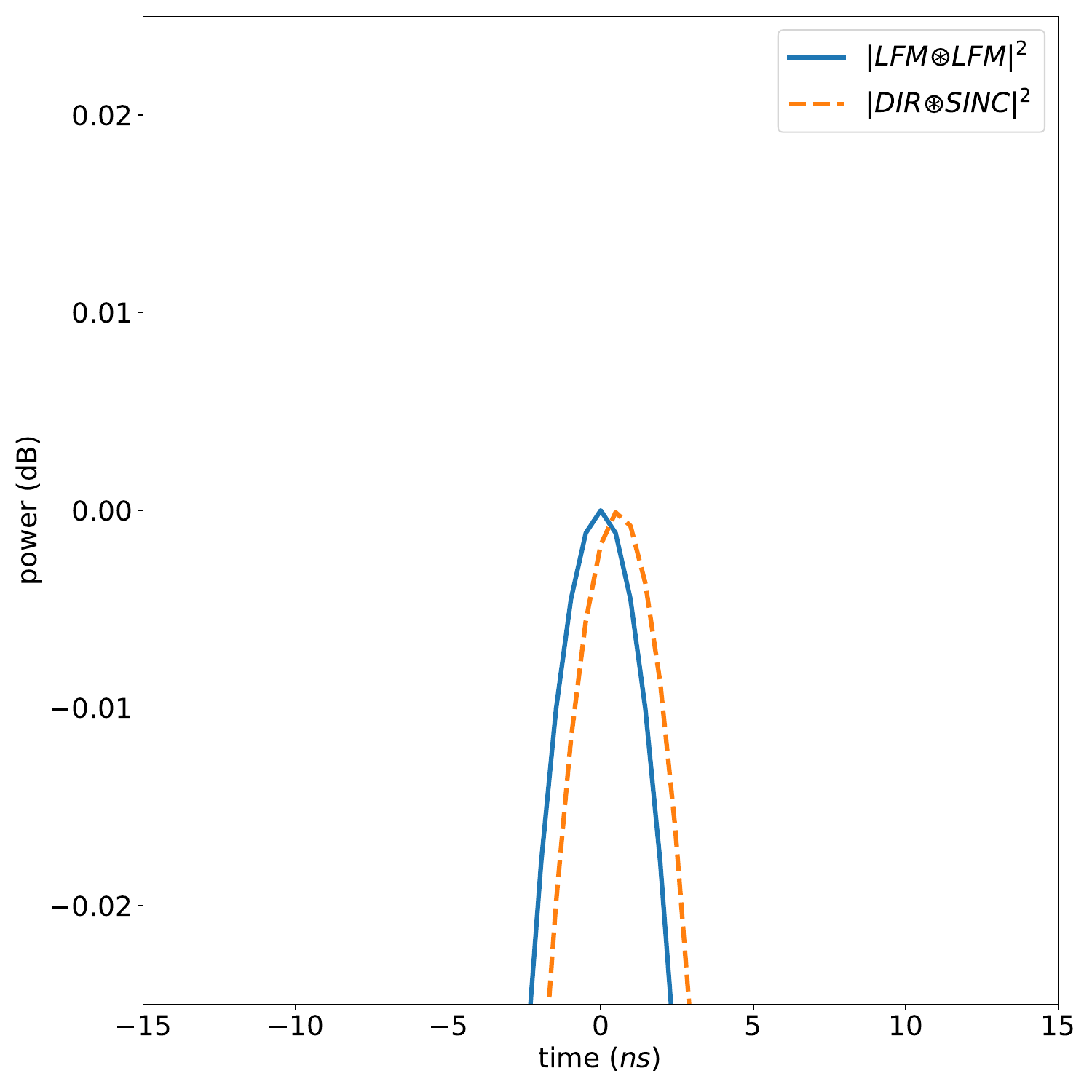}
    \caption{Comparison of windowed Whittaker-Shannon interpolation with the exact analytic interpolation method. This image shows the SNR loss from using sinc interpolation on an LFM waveform with a pulse width $T=0.1~\text{ms}$, sample rate of $f_{s}=2.048~\text{GHz}$, a center frequency of $f_{c}=422~\text{MHz}$, bandwidth of $B=18~\text{MHz}$, and a target velocity of $v=1~\text{km}/\text{s}$. Here the DIR curve represents the analytic solution for Doppler dispersion correction via direct resampling and SINC represents the result of Doppler compensating via windowed Whittaker-Shannon interpolation with a window size of $N_{window}=41~\text{samples}$. The resultant difference in SNR is about a $10^{-4}~\text{dB}$ SNR loss.}
    \label{fig:sincinterpolationsnrloss}
\end{figure}

Whittaker-Shannon interpolation requires a new FIR filter be calculated at every time-step. For each sample, a local filter is calculated using the interpolation shift at the point. This filter will always be whole sample steps plus a fine adjustment when the sampling is between points. Using analytic resampling as a control case for comparison of the windowed Whittaker-Shannon interpolation method, a comparison of the accuracy of sinc interpolation to an exact solution can be seen in Fig.~\ref{fig:sincinterpolationsnrloss}.

\begin{figure}[h]
    \centering
    \includegraphics[width=1\linewidth]{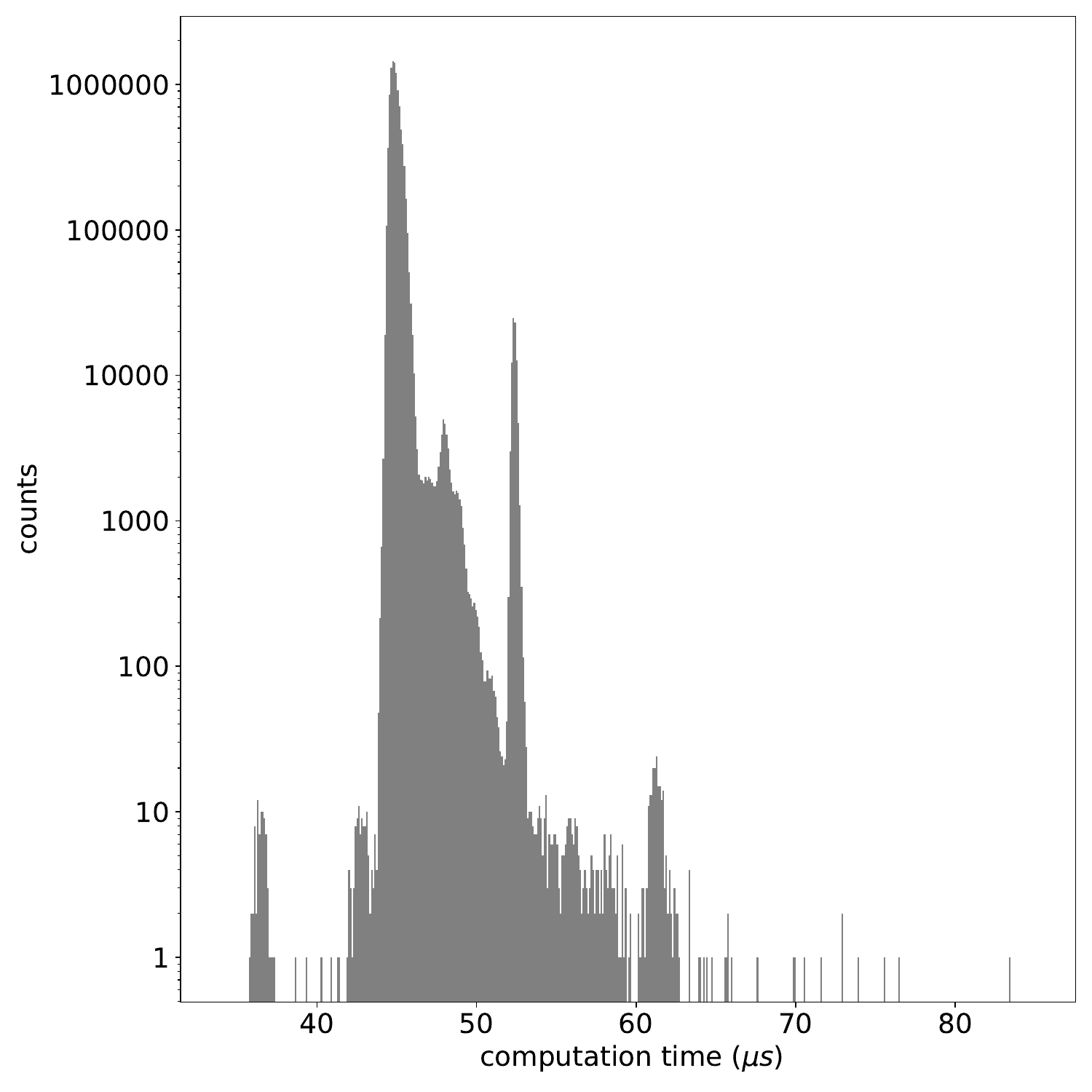}
    \caption{Benchmarks of Whittaker-Shannon interpolation times on an NVIDIA H100 GPU for $10^{7}$ time trials. The input waveform contained $N=2^{19}~\text{samples}$ total with a window size of $N_{window}=25~\text{samples}$. Interpolation was achieved at an average of $45~\mu s$, or a rate of about $11.7~\text{Gigasamples}/s$.}
    \label{fig:sincinterpolationbenchmark}
\end{figure}

Considerations for efficient implementation of sinc interpolation are presented in Appendix~\ref{app:sincoptimizations}. Our implementation of the code resulted in interpolation rates of about $11.7~\text{Gigasamples}/\text{s}$, and a set of benchmarking data can be seen in Fig.~\ref{fig:sincinterpolationbenchmark}. This interpolation only has to be applied for the matched filter. For a radar transmitting at 10\% duty factor, this would mean a rate of $117~\text{Gigasamples}/\text{s}$ could be sampled on receive.

\section{Conclusion}

In this paper, we propose an ionospheric correction algorithm for digital processing with SDR, presented in (\ref{eq:ionospherecorrectionmethod}). Our algorithm is computed directly from the ionospheric delay shown in (\ref{eq:ionosphericdelayequation}). The primary benefits of our method are that it requires no previous information of the input waveform to compute, is valid for both pre-distortion and post-distortion, is valid for arbitrary waveforms, and is straight-forward to implement. Shown in Fig.~\ref{fig:pseudochebyshevcomparison}, we compare our proposed method to a common approximate solution for application on LFMs as well as with an exact analytic method we derived and is shown in (\ref{eq:frequencycubicequationresult}). In doing so, we demonstrate an SNR loss within the error tolerance of the known approximation method, which indicates satisfactory accuracy. In Figs.~\ref{fig:ionobenchmarkwcompression} and~\ref{fig:ionobenchmarkwocompression}, we provide benchmarks of an implementation of our algorithm on an NVIDIA H100 GPU. These benchmarks demonstrate that our method can be viable for application in modern digital radar given appropriate selection of hardware and parallelization.

We also presented a comparison of various Doppler dispersion correction methods via time-dilation interpolation. We highlighted the method of windowed Whittaker-Shannon interpolation as a viable candidate for Doppler dispersion correction in SDRs. Figures~\ref{fig:sincinterpolationsnrloss} and~\ref{fig:sincinterpolationbenchmark} present examples of the accuracy and data processing rates respectively of sinc interpolation in software. We conclude that windowed Whittaker-Shannon interpolation is a viable candidate for applications in real-time DSP.

\bibliography{biblio1}{}
\bibliographystyle{IEEEtran}

\begin{appendices}

\section{Ionospheric Effects on Radio Signals}\label{app:ionospherereview}
Ionospheric effects can be modeled as temporal shifts of individual frequencies of a signal. According to Davies~\cite{davies1990ionospheric} the ionospheric index of refraction, $n_G$, is represented as shown in (\ref{eq:ionosphericrefractionindex}):

\begin{equation}
    n_{G}= 1 + \frac{e^2}{8\pi^{2} m_{e} \epsilon_{0} f^2} N_e.
    \label{eq:ionosphericrefractionindex}
\end{equation}

Here $N_e$ is the ionospheric electron density, $\epsilon_{0}$ is permittivity of free space, and $e$ and $m_{e}$ are the charge and mass of an electron, respectively. Thus, the group velocity of an ionospheric radio wave, $V_G$, is given as (\ref{eq:ionosphericgroupvelocity}):

\begin{equation}
    \begin{split}
    V_{G} = \frac{c}{n_G} & = c\left(1 + \frac{e^2}{8\pi^{2} m_{e} \epsilon_{0} f^2} N_e \right)^{-1}\\
    & = c\left(1 - \frac{e^2}{8\pi^{2} m_{e} \epsilon_{0} f^2} N_e \right) + \mathcal{O}\left( f^{-4} \right).
    \end{split}
    \label{eq:ionosphericgroupvelocity}
\end{equation}

We can truncate the Taylor series expansion in (\ref{eq:ionosphericgroupvelocity}) at $\mathcal{O}\left( f^{-2} \right)$ for sufficiently high values of $f$. To provide context, in SI units this truncation approximation is typically valid for all waveforms such that $f>>6~\text{MHz}$ in Earth's atmosphere. We can use the group velocity to integrate in time and determine the one-way separation $R_e$ between the radar and the target as

\begin{equation}
    R_{e} = \int_{\Delta t} V_{G} dt = c\Delta t - \frac{e^2}{8\pi^{2} m_{e} \epsilon_{0} f^2} \int_{\Delta t} cN_{e} dt.
\end{equation}

Making the substitution $ds=cdt$ to integrate along the propagation path of the pulse yields

\begin{equation}
    R_{e} = c\Delta t - \frac{e^2}{8\pi^{2} m_{e} \epsilon_{0} f^2} \int_{s} N_{e} ds.
\end{equation}

The propagation time given no ionospheric distortion effect, $\Delta t$, would produce a range of $R_{0} = c\Delta t$. We solve for the one-way temporal shift $\tau$ of an ionosphere distorted waveform to be

\begin{equation}
    \begin{split}
    \tau(f) &= \frac{R_{0}-R_{e}}{c} = \frac{e^2}{8\pi^{2} m_{e} \epsilon_{0} cf^2} \int_{s} N_{e} ds\\
    & = \frac{e^2}{8\pi^{2} m_{e} \epsilon_{0} cf^2} E = \frac{K_{2}}{cf^{2}}.
    \end{split}
    \label{eq:ionosphericdelayequationappendix}
\end{equation}

\section{Correction to Pseudo-Chebyshev Polynomial Expansion Method}\label{app:k2correction}

The original Pseudo-Chebyshev method presented in Ref.~\cite{halpin1990propagation} slightly differed from the coefficients we presented in (\ref{eq:T1}) and (\ref{eq:T2}). The original forms of $T_{1}$ and $T_{2}$ were given as:

\begin{equation}
    \begin{split}
    T_{1} = -\frac{aT}{2} &+ \frac{4K_2}{c} \bigg[ \left(f_{0} + \frac{B}{2}\right)^{-2} \\
    & - \left(f_{0} + \frac{B(1-a)}{2} \right)^{-2} \bigg],
    \end{split}
    \label{eq:T1_original}
\end{equation}

and

\begin{equation} 
    \begin{split}
    T_{2} = \frac{4K_2}{c} & \bigg[2\left(f_{0} + \frac{B}{2} \right)^{-2} - \left(f_{0} + \frac{B(1-a)}{2} \right)^{-2}\\
    & - \left(f_{0} + \frac{B(1+a)}{2} \right)^{-2} \bigg].
    \end{split}
    \label{eq:T2_original}
\end{equation}

The values of $T_{1}$ and $T_{2}$ shown in (\ref{eq:T1_original}) and (\ref{eq:T2_original}) only differ by using a factor of $4$ instead of $2$ with each $K_{2}$ term. We originally noticed this difference numerically by comparing parameters for the FFT ionospheric correction method presented in (\ref{eq:ionospherecorrectionmethod}). We then repeated the original derivation from Ref.~\cite{halpin1990propagation} as well as analytically derived an exact solution to (\ref{eq:halpintimedependenceinversion}), which we show in (\ref{eq:frequencycubicequationresult}). We note that in Ref.~\cite{halpin1990propagation} a substitution of a factor of $2$ is made for a factor of $4$, about which the authors state was determined ``by simulation''. Without more information, we are unable to repeat their simulations to determine the validity of that substitution. However, when we rederived the Pseudo-Chebyshev method polynomial expansion, we arrived at the solutions shown in (\ref{eq:T1}) and (\ref{eq:T2}), which the FFT ionospheric correction method agrees with.

\begin{figure}[h]
    \centering
    \includegraphics[width=1\linewidth]{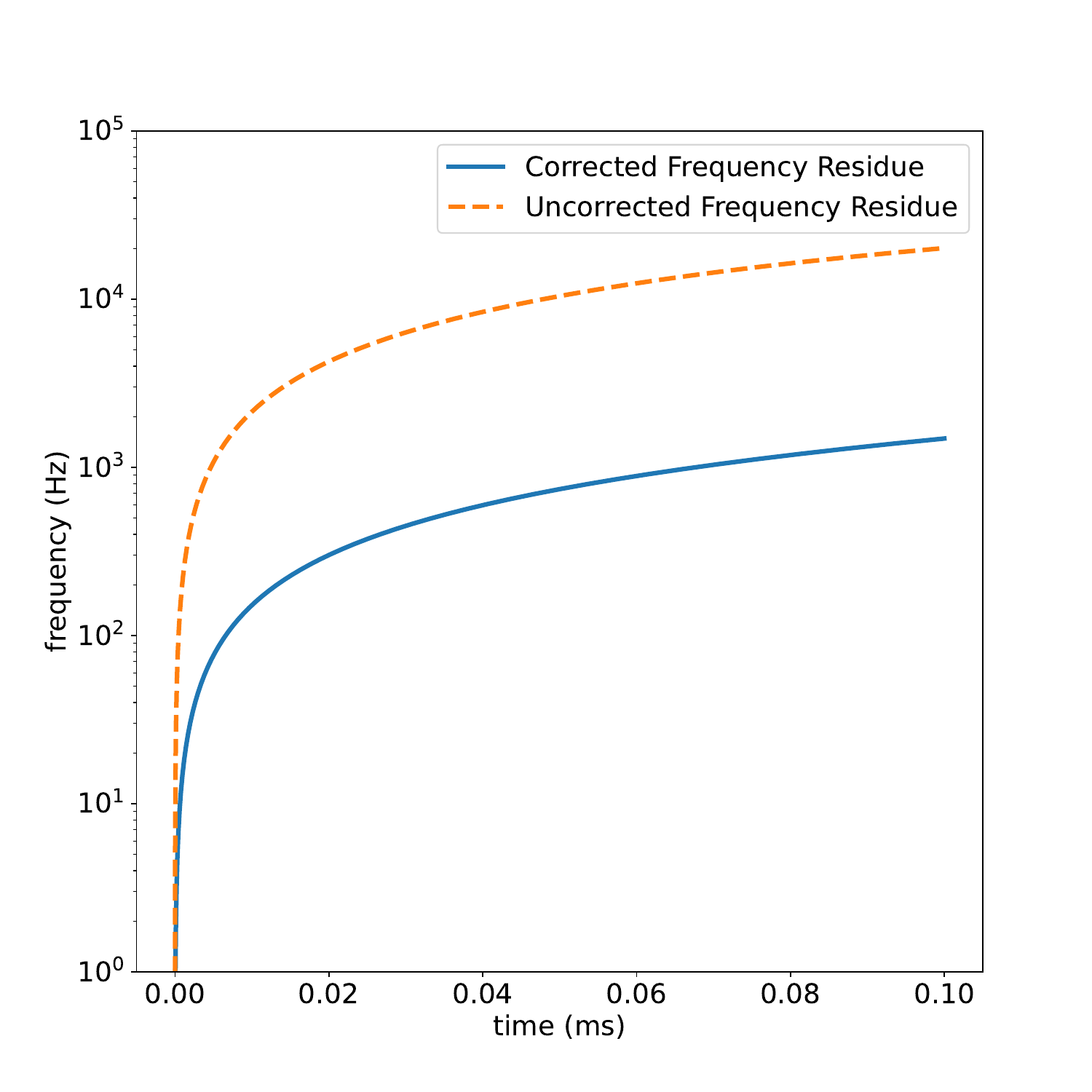}
    \caption{Frequency residuals of the Pseudo-Chebyshev method with our exact solution shown in (\ref{eq:frequencycubicequationresult}) for a given LFM with $B=20~\text{MHz}$, $f_{c}=400~\text{MHz}$, and $T=0.1~\text{ms}$. Here, the corrected residual is the difference returned when using the coefficients in (\ref{eq:T1}) and (\ref{eq:T2}); the uncorrected curve using the coefficients in (\ref{eq:T1_original}) and (\ref{eq:T2_original}). Note that in this case the corrected Pseudo-Chebyshev method resulted in the lower error by over a full order of magnitude.}
    \label{fig:pcm_residules}
\end{figure}

To further verify this change, we compared the original formulation of the Pseudo-Chebyshev method with our corrected coefficients, which is shown in Fig.~\ref{fig:pcm_residules}. As shown in the plot, our corrected coefficient values applied for the given frequency resulted in a difference over a full order of magnitude lower than with the uncorrected method. This difference resulted in about $1~\text{dB}$ of SNR loss when cross correlated with our cubic solution. Based upon the errors from our exact analytic solution, comparison with our FFT-based ionospheric correction method, and analytic derivations, we conclude that the original factor of 2 difference presented in Ref.~\cite{halpin1990propagation} was a slight error, which we have corrected here.

\section{Algorithmic Optimizations for Sinc Interpolation on a GPU}\label{app:sincoptimizations}

For optimal performance, we elected to implement windowed Whittaker-Shannon interpolation on an NVIDIA H100 GPU connected to a $3.25~\text{GHz}$ CPU via PCIe 5.0. Our implementation was written in CUDA C++. Implementation on a GPU allowed us to parallelize a large portion of the computations. Here we will discuss three potential optimizations that were considered, their effect on performance in the H100, and considerations for use on different hardware.

\begin{figure*}[t]
    \centering
    \subfloat[]{
    \begin{minipage}[t]{0.24\textwidth}
        \centering
        \includegraphics[width=1\linewidth]{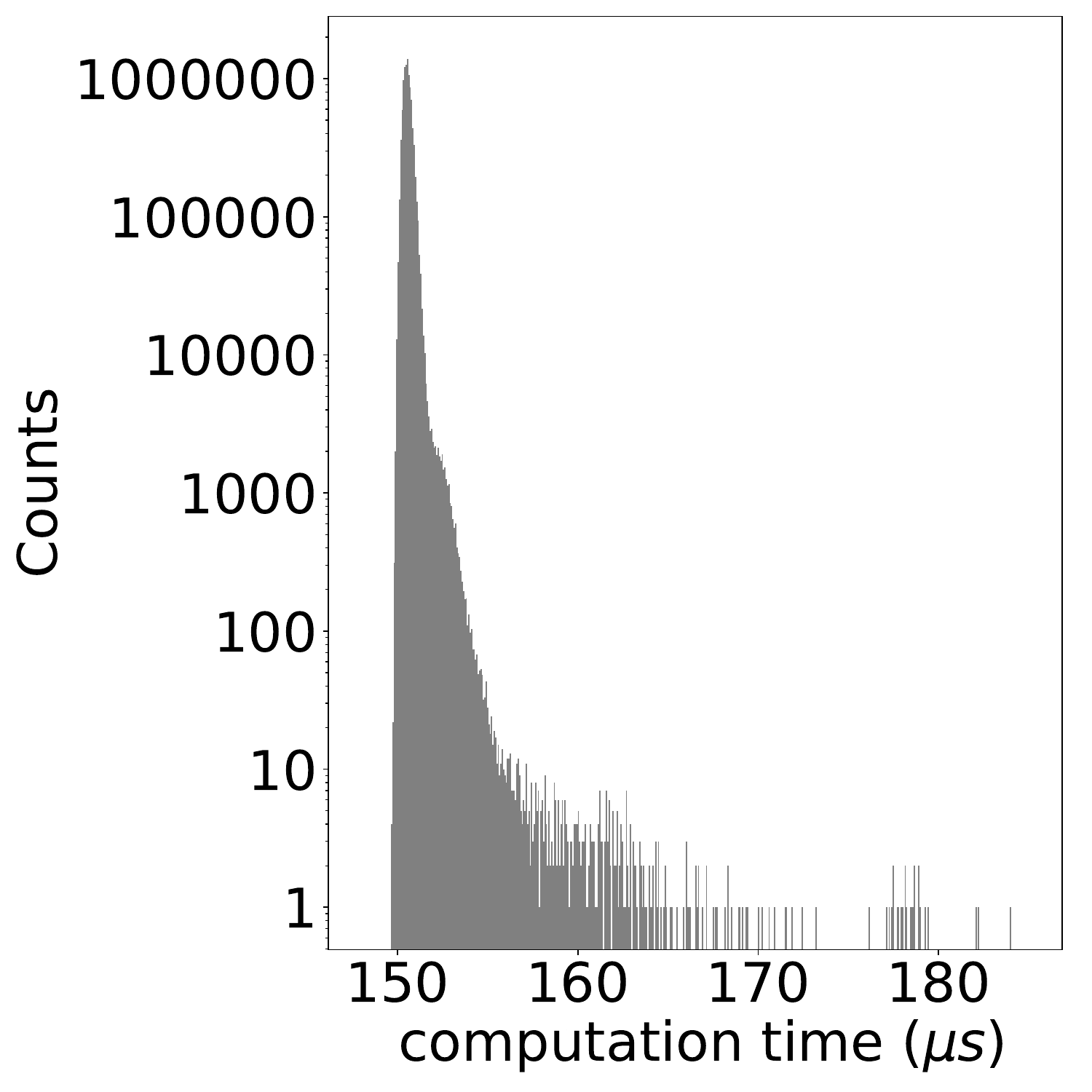}
        \label{fig:sincopparallel}
    \end{minipage}}
    \hfill
    \subfloat[]{
    \begin{minipage}[t]{0.24\textwidth}
        \centering
        \includegraphics[width=1\linewidth]{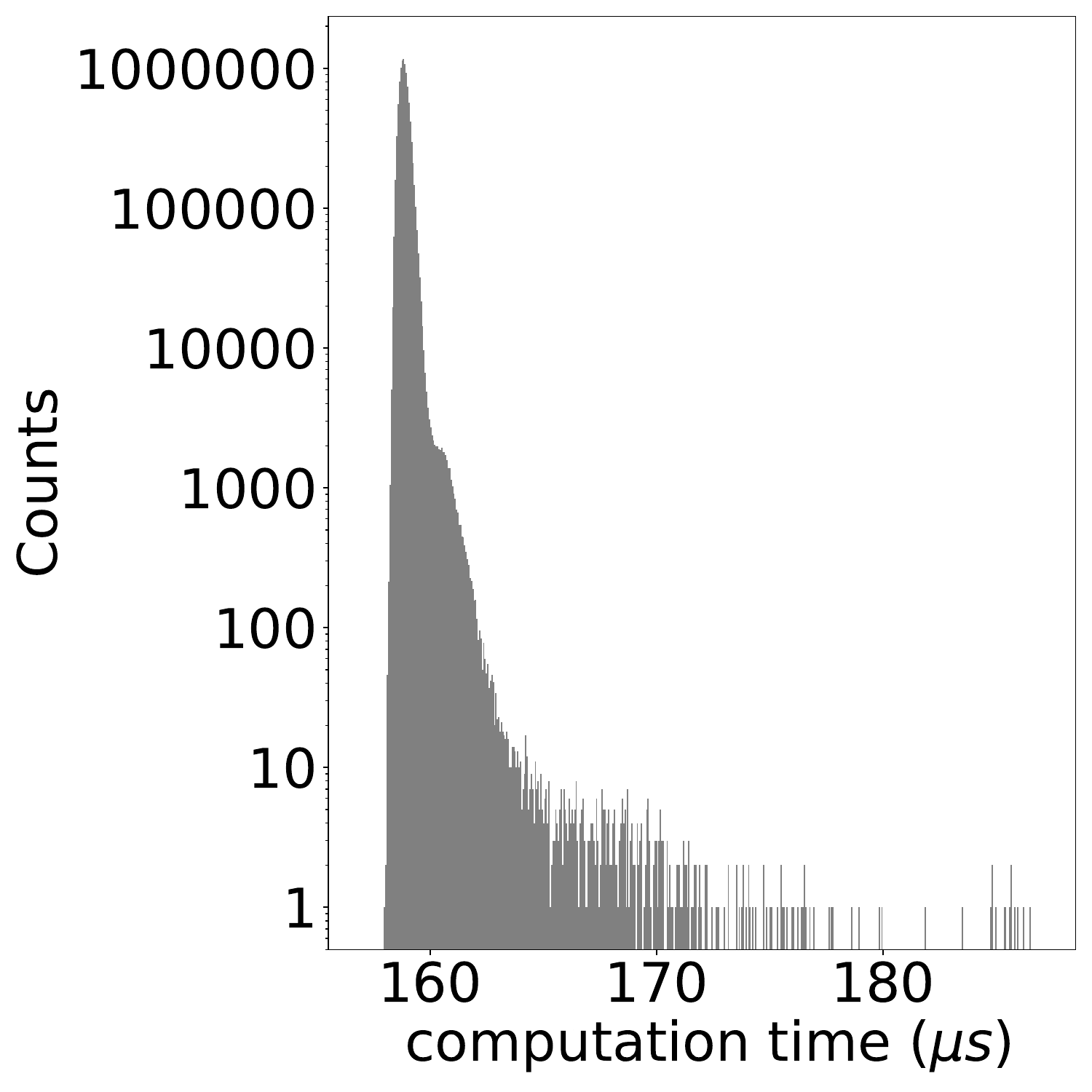}
        \label{fig:sincopshmem}
    \end{minipage}}
    \hfill
    \subfloat[]{
    \begin{minipage}[t]{0.24\textwidth}
        \centering
        \includegraphics[width=1\linewidth]{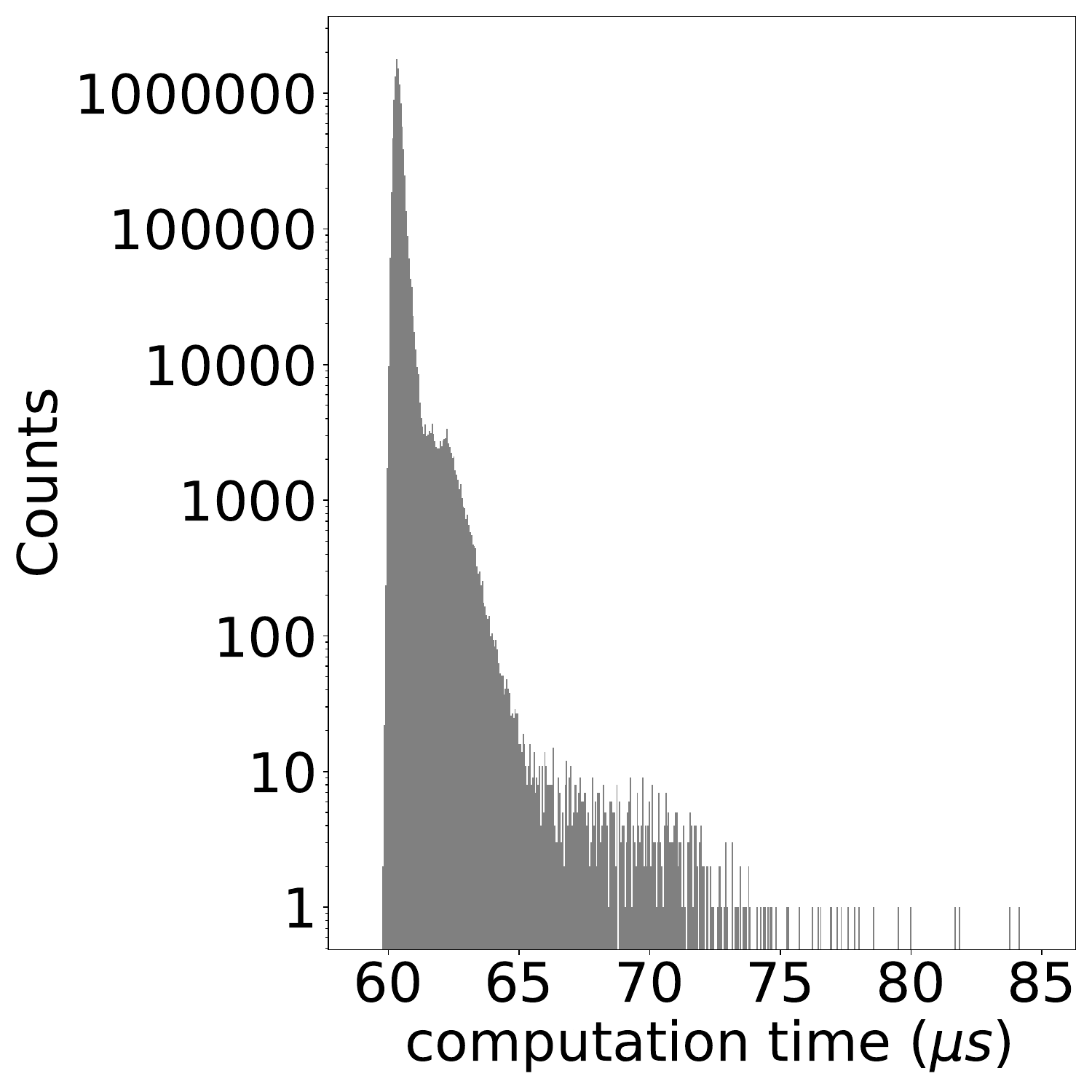}
        \label{fig:sincoplut3000}
    \end{minipage}}
    \hfill
    \subfloat[]{
    \begin{minipage}[t]{0.24\textwidth}
        \centering
        \includegraphics[width=1\linewidth]{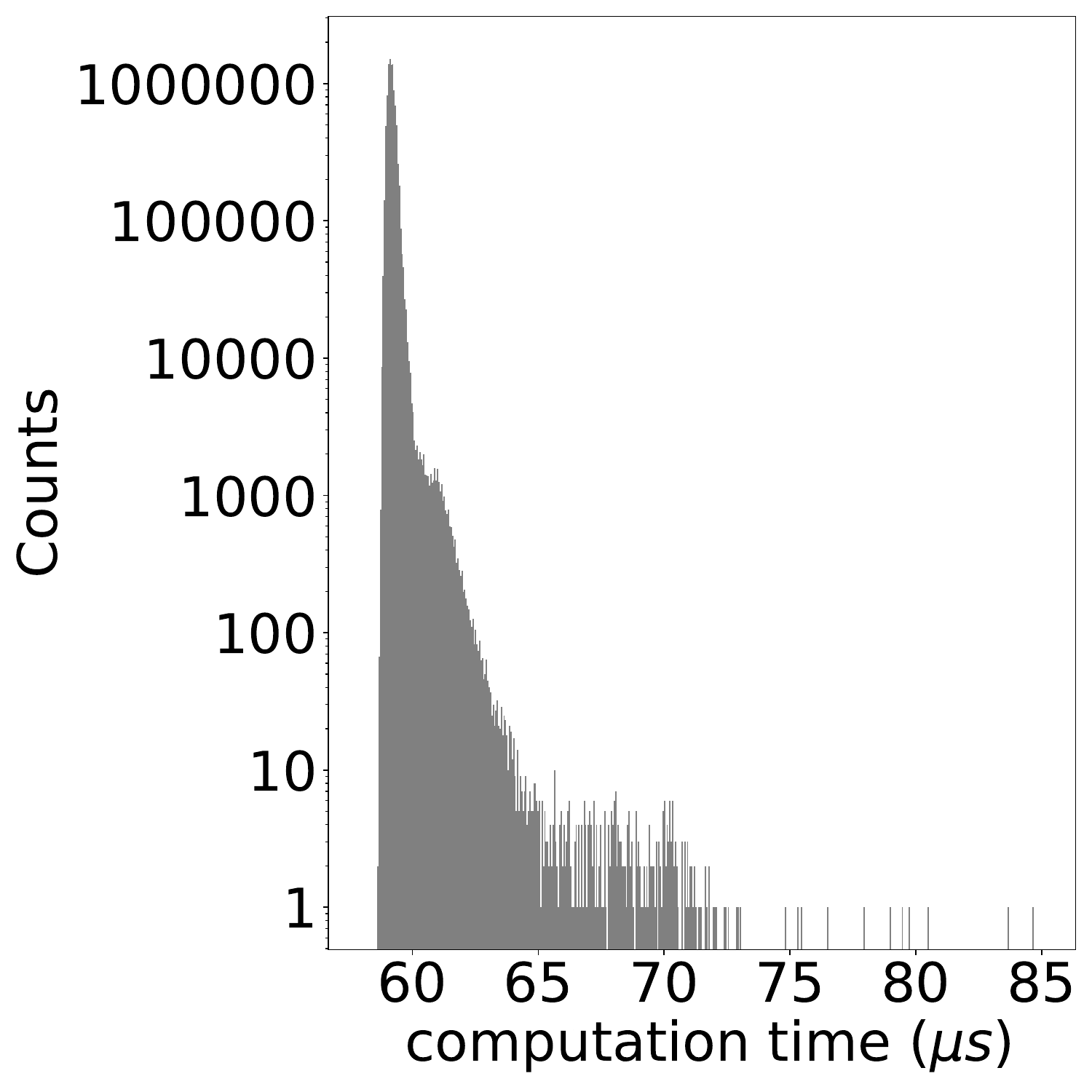}
        \label{fig:sincoplut300}
    \end{minipage}}
    \caption{Benchmarks for this paper's implementation of windowed Whittaker-Shannon interpolation times on an NVIDIA H100 GPU utilizing 32 blocks per kernel. Each plot demonstrates kernel completion times for four different implementations considered and ran for $10^{7}$ time trials. The input waveform contained $N=2^{19}~\text{samples}$ with a window size of $N_{window}=25~\text{samples}$. Figure~\ref{fig:sincopparallel} shows the interpolation time with no optimizations made outside of basic parallelization, which averaged completion around $150.5~\mu\text{s}$. Figure~\ref{fig:sincopshmem} displays the timing data if we first pre-load the input array into shared memory (tiling), which averaged completion around $159~\mu\text{s}$. Figures~\ref{fig:sincoplut3000} and~\ref{fig:sincoplut300} present the benchmarking when a LUT was included. The LUTs were size 3000 and 300 for the Figures~\ref{fig:sincoplut3000} and~\ref{fig:sincoplut300} respectively. These kernels averaged completion in about $60.5~\mu s$ for the LUT size of 3000 and $59.3~\mu\text{s}$ for the LUT size of 300.}
    \label{fig:sincoptimizations}
\end{figure*}

\begin{figure*}[t]
    \centering
    \includegraphics[clip, trim=05em 23em 03em 19em, width=1\linewidth]{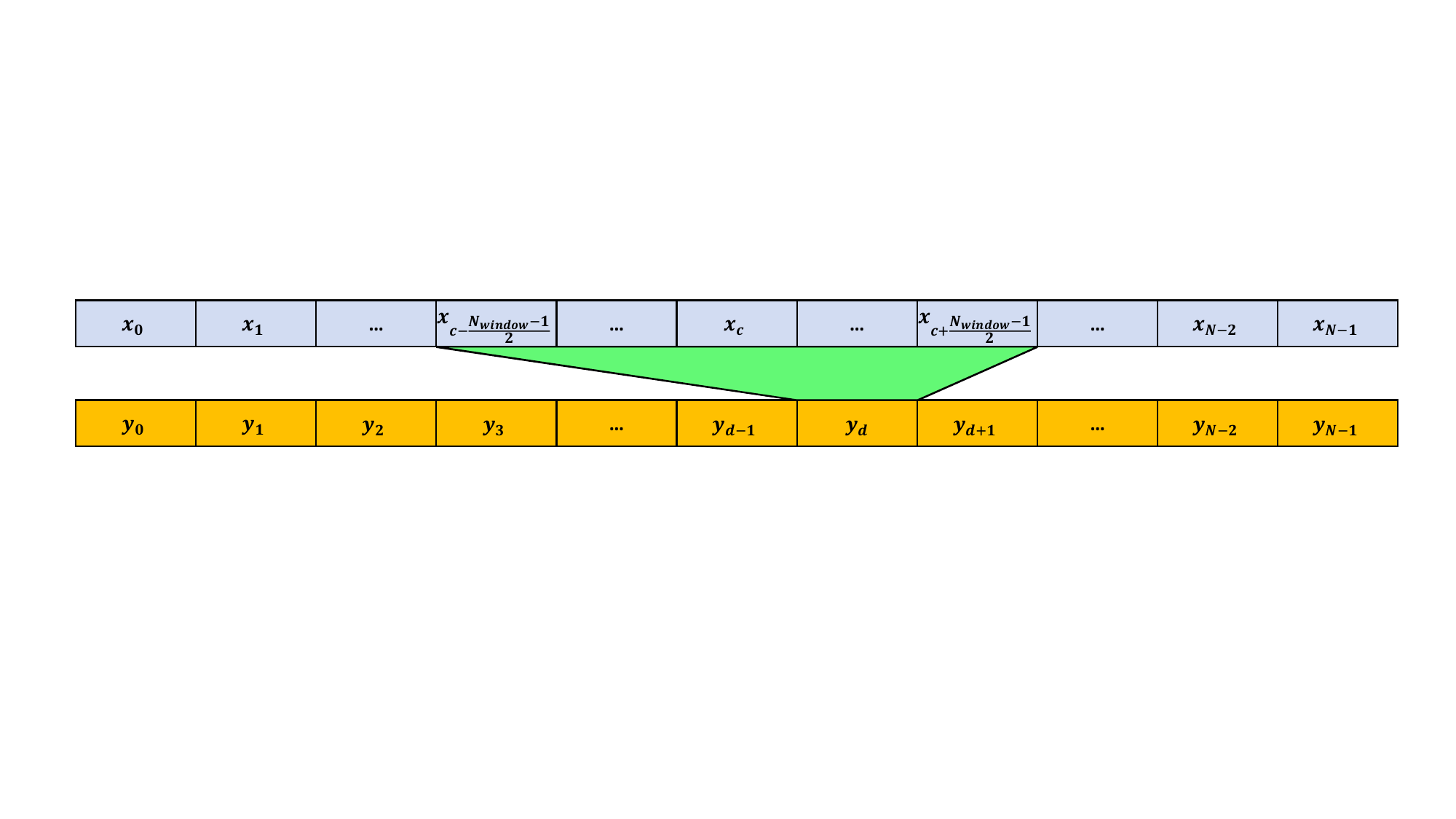}
    \caption{A visual representation of the data accesses during a summation of our parallelized implementation of windowed Whittaker-Shannon interpolation. In this parallelization, every thread in the GPU kernel is allocated to an output index $y_{n}$. For a given thread such that $n=d$, the output index will be $y_{d}$. That worker thread identifies the nearest samples to it in the original waveform, shown as $x_{c}$, which will be the center of the window. The thread then considered all values of $x_{n}$ in the window, which range from $x_{c-(N_{window}-1)/2}$ to $x_{c+(N_{window}-1)/2}$. After the summation, the total is written to the output index.}
    \label{fig:sincparallelization}
\end{figure*}

The primary optimization made during the implementation of this code was parallelization on a GPU. The standard method of parallelizing code on a GPU is to assign each executing thread to an index in the output array. This procedure is not only an efficient way to organize the workload on any given thread, but also assigning each thread a unique output memory address guarantees thread safety during the kernel call. In the case of windowed Whittaker-Shannon interpolation, each thread is responsible for the single summation of a window and writing the resultant interpolant to the output array. Recall from Table~\ref{tab:dopplercomparison} that for an output array of $N$ elements, and a window size of $N_{window}$ elements, windowed sinc interpolation requires $\mathcal{O}\left(N\times N_{window}\right)$ time to complete. If the number of threads in the kernel meets or exceeds $N$, then the parallelized computation on a GPU reduces to $\mathcal{O}\left(N_{window}\right)$ time. For clarity, pseudo-code of this implementation can be seen in Algorithm~\ref{alg:parallelizedsinc} and we visually represented this parallelization in Fig.~\ref{fig:sincparallelization}.\\

\begin{algorithm}
\caption{Parallelized Windowed Whittaker-Shannon Interpolation}\label{alg:parallelizedsinc}
\begin{algorithmic}[1]
    \Procedure{Sinc Interpolation Kernel}{}
    \State{$\text{index} = \text{index of thread in grid}$}
    \State{$\text{stride} = \text{total threads in grid}$}
    \For{$\text{index} < \text{number of samples}$}
    \State{$\text{center} = \text{nearest uninterpolated point to index}$}
    \State{$\text{sum} = 0$}
    \For{input samples in window around center}
    \State{$\text{sinc value} = \text{calculated sinc value}$}
    \State{$\text{sum} = \text{sum} + \text{input sample} \times \text{sinc value}$}
    \EndFor{}
    \State{$\text{output array}[\text{index}] = \text{sum}$}
    \State{$\text{index} = \text{index} + \text{stride}$}
    \EndFor{}
    \EndProcedure{}
\end{algorithmic}
\end{algorithm}

After parallelization on the GPU, the highest-yielding optimization for our implementation of windowed sinc interpolation was to reduce the time cost to compute the values of the $\sinc$ function. The sine function (and therefore the $\sinc$ function) is generally approximated via a Taylor series expansion, which is slow to compute. One can save compute time by precalculating the values of the sinc function for use in the summation and storing them in an array known as a lookup table (LUT). For fast memory access that is visible to all threads in a block, we store the LUT in shared memory on a GPU. The accuracy of the sinc values from the LUT will depend upon the rate at which the $\sinc$ function was sampled, with higher sample rates leading to more accuracy and larger LUTs. The compute time required to initialize this LUT into shared memory depends on the LUT size. Note that since $\sinc$ is an even function, only the positive values of $\sinc$ are required, which saves LUT population time and shared memory. On a GPU, creating a LUT in shared memory offers a significant performance increase during the summation and only comes at the cost of minor wait times to synchronize the threads after populating the LUT. In our case, addition of a LUT improved performance by as much as $2.5\times$ faster interpolation rates when using a LUT size that provided near-exact accuracy. We present pseudo-code for this optimization in Algorithm~\ref{alg:lookuptablesinc}.

\begin{algorithm}
    \caption{Parallelized Windowed Whittaker-Shannon Interpolation w/ Lookup Table}\label{alg:lookuptablesinc}
    \begin{algorithmic}[1]
        \Procedure{Sinc Interpolation Kernel}{}
        \State{$\text{index} = \text{index of thread in grid}$}
        \State{$\text{stride} = \text{total threads in grid}$}
        \State{$\text{sinc LUT} = \text{allocated shared memory}$}
        \For{element in sinc LUT}
        \State{$\text{sinc LUT}[\text{element}] = \text{precalculated sinc value}$}
        \EndFor{}
        \State{synchronize threads}
        \For{$\text{index} < \text{number of samples}$}
        \State{$\text{center} = \text{nearest uninterpolated point to index}$}
        \State{$\text{sum} = 0$}
        \For{input samples in window around center}
        \State{$\text{sinc value} = \text{value in sinc LUT}$}
        \State{$\text{sum} = \text{sum} + \text{input sample} \times \text{sinc value}$}
        \EndFor{}
        \State{$\text{output array}[\text{index}] = \text{sum}$}
        \State{$\text{index} = \text{index} + \text{stride}$}
        \EndFor{}
        \EndProcedure{}
    \end{algorithmic}
    \end{algorithm}

The final optimization we consider here is reduction in the number of read accesses to global memory from the input array of samples. Each thread in the kernel is responsible for $N_{window}$ global memory accesses---one for each element in the summation. Fortunately all threads within a block read from a highly-overlapping region of the input array. Therefore, global memory access can be reduced by allocating a region of shared memory for each block in the kernel, and then having each thread perform a single global memory read to initialize the input array into shared memory. This is a method of preloading the global memory into the block's shared memory, referred to as ``tiling''. Typically tiling is utilized to accelerate memory accesses in a 2D array (with each 2D region being a tile), but it can be applied in our 1D case. The major drawback of the use of tiling is the time cost of thread synchronization that must occur each time the shared memory is updated with a new region of input samples. We found that for our typical choice of interpolation parameters, tiling resulted in about a $9-10~\mu\text{s}$ slow-down in computational time. Because the number of memory accesses to the input array scales as $\mathcal{O}\left( N_{window} \right)$, tiling should become optimal for large enough choice of $N_{window}$, and may be optimal on another choice of hardware where global memory accesses are slower. Because tiling was slightly slower for our use case and for simplicity, we chose to exclude this potential optimization from our implementation of Whittaker-Shannon interpolation. Algorithm~\ref{alg:sharedpreloadingsinc} presents pseudo-code for our version of sinc interpolation with tiling.

\begin{algorithm}
    \caption{Parallelized Windowed Whittaker-Shannon Interpolation w/ Tiling}\label{alg:sharedpreloadingsinc}
    \begin{algorithmic}[1]
        \Procedure{Sinc Interpolation Kernel}{}
        \State{$\text{index} = \text{index of thread in grid}$}
        \State{$\text{stride} = \text{total threads in grid}$}
        \State{$\text{preloaded values} = \text{allocated shared memory}$}
        \For{$\text{index} < \text{number of samples}$}
        \For{element in preloaded values}
        \State{$\text{preloaded values}[\text{element}] = \text{global value}$}
        \EndFor{}
        \State{synchronize threads}
        \State{$\text{center} = \text{nearest uninterpolated point to index}$}
        \State{$\text{sum} = 0$}
        \For{input samples in window around center}
        \State{$\text{input value} = \text{value from preloaded values}$}
        \State{$\text{sinc value} = \text{calculated sinc value}$}
        \State{$\text{sum} = \text{sum} + \text{input sample} \times \text{sinc value}$}
        \EndFor{}
        \State{$\text{output array}[\text{index}] = \text{sum}$}
        \State{$\text{index} = \text{index} + \text{stride}$}
        \State{synchronize threads}
        \EndFor{}
        \EndProcedure{}
    \end{algorithmic}
    \end{algorithm}

Note that the implementations of a LUT and tiling both require shared memory allocation that is related to $N_{window}$ in size. It is possible that tiling and the use of a LUT will be incompatible on devices with sufficiently limited shared memory and/or a sufficiently large choice of $N_{window}$. Based on our benchmarks and because the number of sinc value calculations scales as $\mathcal{O}\left( N_{window} \right)$, it seems likely that the implementation of a LUT will most-likely be optimal in any situation where a LUT and tiling are mutually exclusive.

Most further optimizations for a GPU implementation of sinc interpolation are much more specific to the given hardware used, such as maximal occupancy requirements. These optimizations are more sophisticated than we intend to include here. A comparison of the efficiency of each considered optimization on our hardware limited to 32 blocks per kernel is shown in Fig.~\ref{fig:sincoptimizations}.

For completeness, we note that a great deal of work has been done on efficient Whittaker-Shannon interpolation by padding elements in the frequency domain via the use of DFTs~\cite{olkkonen1990fast, yaroslavsky2002fast}. For efficient processing, these methods require that the number of interpolants be proportional to the number of input samples by a power of 2. This condition is compatible with applications where the number of interpolants, $N_{I}$, is selectable such as image processing. When this condition is met, Whittaker-Shannon interpolation is achievable with time complexities of $\mathcal{O}\left(N_{I}\log{N_{I}}\right)$. Interpolation for Doppler dispersion correction requires a specific choice of $N_{I}$ that depends upon $N$ and the target velocity. The variable number of interpolants results in variable time complexity to complete the required FFTs. For our use case, these methods resulted in an unpredictable signal processing rate that was consistently slower than all implementations shown in Fig.~\ref{fig:sincoptimizations}---in some cases by a factor of $10\times$ slower rates. These trade offs are extremely similar to considerations for the FFT P/Q resampling method. We conclude that for radar signal processing, performing the interpolation in the time domain directly from (\ref{eq:whittakershannoninterpolation}) is often more efficient and always more predictable in terms of kernel completion times.

\end{appendices}

\end{document}